\documentstyle[12pt]{article}

\begin{document}
\title{Suppression of Hadroproduction in Nuclei\thanks{Work
funded in part by the Russian Foundation of Fundamental Research (contract
No 93-02-14381) and by the Deutsche Forchungsgemeinschaft (SFB 201) }}
\author{S.V.Akulinichev\thanks{E-mail address: AKULINICHEV@INUCRES.MSK.SU} \\
{  \it Institute for Nuclear  Research, 60-th October Anniversary}\\
{  \it  Prospect 7a, Moscow 117312, Russia}\\
 and\\
{  \it Institut f\"{u}r Kernphysik, Universit\"{a}t Mainz, 6500 Mainz,
Germany}}
\maketitle
\begin{abstract}
We argue that nuclei are not transparent for fast projectile
partons. Color transparency is effective
for final state interactions in heavy particle production,
though nuclear filtering of initial partons
can reduce the cross sections.
We show that short-ranged initial state interactions, which have been neglected
so far, can be  important  in hadroproduction.
With the present scenario of hadroproduction a qualitative discription of data
can be obtained.

\end{abstract}
\newpage
{\bf 1. Introduction}

The data for hadroproduction on nuclei contradict the standard model of
color transparency (CT): the suppression of $J/\psi$- and
$\Upsilon$- production in nuclei\cite{Alde} is far beyond the predictions
of standard CT model.
The observed large distortion of $p_{T}$-distributions in all
hadroproduction reactions in nuclei, especially in Drell Yan (DY)
production, can hardly be described without the contribution of
initial state (IS) interactions. In some papers  the elastic IS
interactions have been taken into account
to explain the $p_{T}$-dependence (see, e.g., Refs.\cite{Pirn,Sal}).
We show that the hadroproduction suppression in nuclei
can be attributed to short-ranged
IS interactions of projectiles, which have been neglected so far.
We argue that these interactions
do not contradict the CT forecasts and that CT is not very effective
for projectile partons.

According to CT forecasts,
the propagation of a color singlet partonic configuration in nuclei
is described by cross sections  vanishing like $r^{2}$,
where $r$ is the configuration transverse size\cite{Gun}.
Though this result was
 first obtained within the Low-Nussinov model,
it has a plausible physical interpretation and got a larger application.
A simple analog is the non-interaction of photons with point-like dipoles.
Heavy quarkonia are first produced as bare singlet quark-antiquark
configurations with $r\sim 1/Q$  and further
hadronize and get a normal hadronic size far outside the nucleus. Therefore
the final state interactions for heavy quarkonium production in nuclei
are unimportant. The situation with IS interactions is
quite different. Many authors argued that only small size projectile
parton configurations are involved in the annihilation
with large $Q^{2}$ (see,e.g., Ref.\cite{Nik}).
In fact, the annihilating (and produced) partons must have the
transverse separation $\sim 1/Q$. But the annihilating partons belong to
different hadrons (projectile and target), while the produced partons
belong to the same hadron (or pre-hadron).
There is no restriction for transverse separation between partons from the same
projectile hadron and the closeness of
annihilating  partons has no effect for the propagation of projectile
hadrons in nuclei.
Therefore large size partonic configurations of projectile hadrons
could contribute to hadroproduction as well,
but these configurations
are filtered out by the IS nuclear filtering and the hadroproduction
on back nucleons is  less effective.
The above argumentation is not  valid for elastic formfactors
since in this case only small partonic configurations
of the size $\sim 1/Q$ from both initial and
final hadrons can contribute. This is so because all
partons must correlate during the time $\sim 1/Q$ in order for the reaction
to be elastic. But only one active parton can participate in inelastic
reactions and other projectile partons have no knowledge
of what happens with that parton.

{\bf 2. Initial state interactions.}

Several authors\cite{Muel1,BBL} demonstrated
the cancellation of IS gluon exchange diagrams in hadroproduction
on nuclei at large $Q^{2}$.
The contribution of exchange gluons is shown either to vanish or, for
colliniar gluons,
to enter the familiar structure functions\cite{CSS,Bodw}.
We now focus on the contribution of active - spectator IS interactions
with sufficiently large exchange momenta (see below).
Assume that the operators $Q_{1}$ and $Q_{2}$ describe the IS interactions
and the final
annihilation of a projectile parton, resp., $|i>$ is the assymptotic initial
state and $|f>$ is the final state after the interaction $Q_{2}$.
We neglect other interactions of the parton.
The vanishing of
Feynman diagrams describing the two-step process, i.e.
the annihilation after the IS scattering, means that
\begin{equation}
< f | Q_{2} Q_{1} | i > = 0.
\end{equation}
For hard IS interactions, this equation is justified because the
transverse momentum of an active parton after the IS interaction
is assumed to be larger than the typical transverse momentum of annihilating
partons\cite{BBL}.
Many authors concluded that the interaction $Q_{1}$ can be neglected
if (1) is valid.
This would be correct if particle fields were infinite
(remember that Feynman
diagrams describe  interactions of infinite fields) and particles could
interact at any space-time point.
In this case we could use the same assymptotic initial state $|i>$ for
both interactions $Q_{1}$ and $Q_{2}$. In reality, particles are localized
objects and assymptotic initial states for  front and
back nucleons can be different. We assume that the interaction $Q_{1}$ is
short-ranged, i.e. for any momentum $l$ involved in
this interaction
\begin{equation}
l \gg L^{-1},
\end{equation}
$L$ is the target length  in the lab. frame.
This equation is our definition of hard interactions.
The interaction $Q_{2}$
is also short-ranged at large $Q^{2}$.
In this case  the de-excitation time (or the coherence length) of the
projectile hadron is
unimportant because this parton participates either in $Q_{1}$ or
in $Q_{2}$ interaction, but not in both interactions due to (1).
The second step of the process is described by
the amplitude $< f | Q_{2} | i' >$, where $| i' >$ is the assymptotic initial
state for the interaction $Q_{2}$,
\begin{equation}
| i' > = C | i > + \sum_{n} | n >\: < n | Q_{1} | i >,
\end{equation}
$ | n > $ are final states after the  interaction $Q_{1}$.
Note that $| i' >$ is the assymptotic state for the interaction $Q_{2}$
only under the condition (2), when the interactions $Q_{1}$ and $Q_{2}$
are independent.
Eq. (1) can now be rewritten as
\begin{equation}
< f | Q_{2} | n > = 0.
\end{equation}
If the states $ | n > $ and $ | i > $
are orthogonal, and this is a reasonable assumption for hard IS
interactions, then the probability conservation demands
\begin{equation}
< i' | i' > = < i | i > = 1 , \:\: C^{2} = 1 - \sum_{n} |\: < n | Q_{1} | i >
\:|^{2}.
\end{equation}
For hard IS interactions, the condition (4) can be fulfilled indeed.
In this case the whole reaction is described  by the amplitude
\begin{equation}
< f | Q_{2} | i' > = C\: < f | Q_{2} | i >.
\end{equation}
The macroscopic factor $C$ describes the probability to escape
hard IS interactions. Note that the interaction $Q_{1}$,
which is responsible for this factor, contains
the lowest powers of $\alpha_{S}$.
For  soft IS interactions with $l\sim L^{-1}$, the result (6) is not
correct because in this case $ |i'>$ is not an assymptotic state for the
interaction $Q_{2}$. In this case the results of Refs.\cite{Muel1}
-\cite{Bodw} for soft gluons should be applied instead.
In Refs.\cite{BBL,Bodw} the following target-length condition for the validity
of factorization was obtained
\begin{equation}
Q^{2} \gg x_{q}LMl_{T}^{2},
\end{equation}
where $x_{q}$ is the projectile parton's
Bjorken variable, $M$ is the target mass and $l_{T}$ is the exchanged gluon
transverse momentum. From the derivation of this condition it follows that
for hard IS interactions, (7) can be interpreted as a
condition for the validity of (4).
In fact, if (7) is fulfilled then the only remaining contribution of IS
interactions to the two-step amplitude comes from colliniar gluons
\cite{CSS,Bodw}.
As we have shown, in this case
the factor $C$ suppresses the
hadroproduction cross sections, provided the target-length condition (2)
is satisfied.

It was argued\cite{CSS} that hard IS interactions cannot
take place because in the c.m. frame projectile and target partons remain
space-like separated until the final annihilation takes place. In fact,
in that frame the nucleus is viewed as a thin pancake due to Lorentz
contraction, whereas the wave-lengths of projectile partons are finite
and larger than in the target rest frame.
At certain energies, the  wave-lengths become larger than the
contracted nucleus length and
the condition (2) is not satisfied in the c.m. frame.
However, the wave-length (and uncertainty principle) argumentation should
be used with more care in fast moving frames.
This can be illustrated by the following example.
Assume that in the target rest frame  a projectile (or exchanged) particle,
which directly couples to target particle fields,
has the four momentum $p = (E,\vec{P})$
and the longitudinal separation between two target particles is $ L$.
In this frame the target system is well described by the instant-form
dynamics and we may say that these two target particles are separated by the
space-time interval $\xi  = (0,0,0,L)$. The invariant macroscopic
target-length condition
can be obtained from the requirement that the
phase accumulated by the fast particle on the interval $\xi$ is sufficiently
large,
\begin{equation}
p\xi \gg 1.
\end{equation}
In this case local interactions of the fast particle with two target
particles will be independent.
In the target rest frame, this condition takes the form
(2), which is non-relativistic with respect to the target.
But it is misleading to use (2) instead of (8) in fast moving frames.
In such frames a composite system cannot be unambigiously described by
the instant form wave functions and the relative position
of two target particles
at the same time is not well defined. For the relativistic description
of composite systems in fast moving frames it is more appropriate to use
the light-cone formalism\cite{Leut}, in which all systems are described at
fixed light-cone "time". In this formalism a transition from one frame
to another, as well as a composite system description in fast moving frames,
is unambigious. Note that there is no Lorentz contraction of a target
light-cone "size" measured at a fixed light-cone "time" in the c.m. frame.
A more detailed discussion of this problem is beyond the scope of this
paper.
{}From the invariant condition (8) it follows that if two target
particles are separated for a projectile in the target rest frame, they
are also separated in any other frame. Otherwise cross sections for
composite targets would not be invariant values. The cross section of,
say, photon scattering from a dipole would be frame-dependent because the
relation between the photon wave-length and the dipole size depends
on the choice of frame. For very energetic photons,
this would lead to the vanishing of photon-dipole
cross section in the c.m. frame due to Lorentz contraction of the
dipole size. In reality, high momentum photons interact almost
independently with dipole charges.

We conclude that there is no theoretical
evidence to ignore hard IS interactions in hadroproduction on nuclei.
Note that our conclusion does not contradict either the factorization
theorem for one-nucleon target, because in this case $|i'>\approx|i>$,
or the "weak" factorization theorem\cite{CSS,Bodw}
for nuclear targets, because the factor $C$ only changes the
normalization of the amplitude and the annihilation vertex
is described by the same factored form.
In Ref.\cite{Gavin} it was suggested that the hadroproduction
suppression should be attributed to energy loss of incident partons in
nuclei.
However, the energy loss needed to explain the data is too large
\cite{Brod3}, if the same active parton is going to annihilate
afterwards.
In contrast to Ref.\cite{Gavin}, we assume that some IS interactions
cannot be followed by the annihilation of the same active parton.
In our conjecture, IS interactions reduce the flux of partons, which
are suitable for hadroproduction.
The contribution of these IS interactions cannot be included in target
or projectile hadron structure functions.
These structure
functions contain diagrams with deep inelastic interactions of active
partons, while  the factor  $C$ does not.

{\bf 3. Numerical results.}

To simulate hard IS interactions, we introduce the phenomenological
parton-nucleon absorption cross section $\sigma_{abs}$.
We assume that these interactions are approximately the same for quarks
and antiquarks, $\sigma_{abs}^{q}\approx \sigma_{abs}^{\overline{q}}$.
It is appropriate to find $\sigma_{abs}^{q}$
from the Drell-Yan (DY) production by
pions. This reaction probes the quark content of a target and
is not affected either by final state
interaction or by the possible excess pion contribution.
The  ratio of nuclear and nucleon cross sections for this reaction,
$R^{\pi}_{DY}(x)$, and the electroproduction cross section ratio,
$R_{EMC}(x)$, are connected by the equation
\begin{equation}
R^{\pi}_{DY}(x) = F\:\:R_{EMC}(x).
\end{equation}
The factor $F$, which is the generalization of the macroscopic factor
$C^{2}$, can be written in the eikonal form
\begin{equation}
F = \frac{1}{A \sigma_{abs}} \int d^{2}b\:(1 - e^{-\sigma_{abs}\:
T(\vec{b})}),
\end{equation}
$T(\vec{b})$ is the profile function (we assumed  $\sigma_{abs}
\gg \sigma_{DY}$).

The data\cite{Bord} show that in accordance with (9) the ratio $R^{\pi}_
{DY}(x)$ is systematically below $R_{EMC}(x)$, whereas their x-dependences
are quite similar (see Fig.1).
The value $\sigma_{abs}^{q}=
2mb$ gives a reasonable fit for the $E_{\pi}$ = 140 GeV data,
while $\sigma_{abs}^{q} \approx$  1.5 mb would better fit the combined
data. Note that the data for $E_{\pi}$ = 286 GeV
have larger error bars and do not demonstrate a smooth behaviour.
In reality, $\sigma_{abs}$ may be $x$- and $p_{T}$- dependent.
Therefore the constant parameter $\sigma_{abs}$
can only reproduce the bulk of the suppression.

It is usually accepted that the main subprocess for  the quarkonium
production is the annihilation of projectile and target gluons.
In this case $\sigma_{abs}^{g}$ describes the gluon-nucleon IS interactions.
In Fig.2 the CT model predictions\cite{Pire}
for the $J/\psi$  and $\Upsilon$
production are compared to the data and we can see that
the standard CT scenario is not good enough.
There were many attempts to explain this discrepancy.
Among them, we can mention two possible explanations: the gluon shadowing in
target nuclei\cite{Satz} and the  final state interaction of produced heavy
quarks\cite{Huf}. There is no final agreement what is the origin of
parton shadowing, which have been observed in deep inelastic lepton-nucleus
scattering at small $x$. Therefore it is not clear whether the same shadowing
should take place at time-like $Q^{2}$. The gluon shadowing cannot entirely
explain the quarkonium production suppression for the following reasons:
1) the $\Upsilon$-production is also suppressed  though the measured $x$-region
in this case\cite{Alde} does not correspond to the shadowing effect;
2) it is hard to explain the different  magnitude of nuclear suppression in
DY production by pions, in DY production by protons, in $J/\psi$-production
and in $\Upsilon$-production. At most, the parton shadowing is not the
only source of hadroproduction suppression.
The additional final state interactions of quarkonia, that are not included
in the standard CT model, can hardly explain the suppression for the
following reasons: 1) the observed suppression of $\psi'$-production is the
same
as the suppression of $\psi$-production, though the transverse size of
valence parton
configuration is much larger in the first case; 2) the suppression of
$J/\psi$-production by photons is not so large\cite{Aubert}, which means
that the suppression is due mainly to IS interactions.

According to our conjecture, the production cross section ratio
for this reaction, $R_{Q}$,can
be written as
\begin{equation}
R_{Q} = F R_{Q}^{CT},
\end{equation}
where $R_{Q}^{CT}$ is the standard CT result for the ratio
and the factor $F$ describes  hard IS interactions of projectile gluons.
As it follows from Fig.2, the best fit of data is with
$\sigma_{abs}^{g}(J/\psi)\approx$
4 mb and $\sigma_{abs}^{g}(\Upsilon)\approx$ 2 mb.

The cross section $\sigma_{abs}$= 1.5 - 4 mb provides
an overall description of hadroproduction data. From the comparison
with the data it follows that 1) $<\sigma_{abs}^{g}>\approx 2 <\sigma_{abs}
^{q}>$ and 2)$\sigma_{abs}^{g}(J/\psi) > \sigma_{abs}^{g}(\Upsilon)$.
The first result can be explained by the color factors which
make the gluon-nucleon cross section twice as large as the quark-nucleon
cross section. The second result could be explained, for example, by the
contribution of non-fusion mechanism of charm production\cite{Brod2}  or by the
contribution of quark-antiquark annihilation subprocess to the bottomonium
production.
Both mechanisms make the IS nuclear effects in
$J/\psi$-production larger than in $\Upsilon$-production.
However, the data for diffractive charm  production\cite{Kod}
do not support the non-fusion mechanism of $J/\psi$-production.
Note that the $J/\psi$-meson is probably too light for the validity of
factorization.
The factorization condition (7), which
can be rewritten as
\begin{equation}
Q^{2}\gg 0.25\:A^{2/3}GeV^{2},
\end{equation}
is fulfilled for $\Upsilon$-production and, partially, for DY
reactions, but not for $J/\psi$-production. This is a possible additional
source of nuclear effects in the last case.
And, finally, the Sudakov suppression of color correlations\cite{Muel1,BBL}
is not very effective for $J/\psi$-production because in this case the Sudakov
formfactor is not so small, $|S(Q^{2})|^{2}\sim 1/3$ at $Q^{2} = 10 GeV^{2}$.
This should also make nuclear effects in $J/\psi$-production
larger than in $\Upsilon$-production.
With all these remarks, the suggested suppression mechanism gives a
satisfactory
phenomenological description of the data.

In the DY production by protons, which probes the antiquark content of
a target,  the measured cross section ratio is close to 1 \cite{Alde1}.
Therefore many authors concluded that there were no excess pions (sea
enhancement) in nuclei. This would be correct in the absence of IS
interactions. In Ref.\cite{Akul} it was shown that the contribution
of a typical amount of excess pions, $\sim 0.1$ pions per nucleon,
is roughly compensated by IS interaction with $\sigma_{abs}^{q}
\approx 2 mb$.
This means that the models for the nuclear EMC-effect, assuming some
pion excess in nuclei, are rather supported by the DY production data
than ruled out as many authors concluded.
An indication of pionic contribution to the
DY production by protons could be the data for the angular distribution
of leptons, which has an important $sin^{2}(\Theta)$-term for $\pi N$
collisions.
Following our conjecture, there is an additional source of the relative
enhancement of hadroproduction with large $p_{T}$:
projectile partons from small size configurations have larger $p_{T}$ than
partons from large size configurations, which are filtered out by IS
interactions. This effect will be considered elsewhere.

We conclude that the CT model can describe the available data for
hadroproduction on nuclei, provided the IS interactions are taken into
account.
The $J/\psi$-production suppression in heavy ion collisions could be
a signal of quark-gluon plasma. Here we considered a new  source of nuclear
suppression  of hadroproduction. A firm conclusion about the observation
of quark-gluon plasma cannot be made without determining all major nuclear
effects in hadroproduction.

I would like to thank the Institute f\"{u}r Kernphysik of Mainz University
for the warm hospitality.

\newpage

{\bf Figure captions.}

Fig.1. The ratio $R^{\pi}_{DY}(x)$, calculated for Fe with
$\sigma_{abs}^{q}$= 0 ($R_{EMC}(x)$, dashed curve), 2mb (solid curve) and
4mb (dotted curve).
The data are for the DY production  by pions
measured on W and D at $E_{\pi}$ = 140 GeV (dots) and $E_{\pi}$ = 286 GeV
(circles) \cite{Bord}.

Fig.2. The ratio $R_{Q}$ for the quarkonium production
calculated with $\sigma_{abs}^{g}$=0 ($R^{CT}_{Q}$, dashed curves),
2mb (solid curves) and 4mb (dotted
curves). Upper curves are for $\Upsilon$ and lower curves for $J/\psi$
production. The data for $\Upsilon$ (dots) and $J/\psi$
(circles) production by protons are from Ref.\cite{Alde}.

\end{document}